\documentclass[twocolumn,showpacs,preprintnumbers,amsmath,amssymb]{revtex4}
\usepackage{graphicx}
\usepackage{dcolumn}
\usepackage{bm}
\begin{document}
\title{Strength Reduction in Electrical and Elastic Networks}
\author{J.S. Espinoza Ortiz,$^1$ Chamith S. Rajapakse,$^1$ and Gemunu H.
Gunaratne$^{1,2}$} \affiliation{$^1$ Department of Physics,
University of Houston, Houston, TX77204} \affiliation{$^{2}$ The
Institute of Fundamental Studies, Kandy 20000,Sri Lanka}
\begin{abstract}
Particular aspects of problems ranging from dielectric breakdown
to metal insulator transition can be studied using electrical or
elastic networks. We present an expression for the mean breakdown
strength of such networks. First, we introduce a method to
evaluate the redistribution of current due to the removal of a
finite number of elements from a hyper-cubic network of
conductances. It is used to determine the reduction of breakdown
strength due to a fracture of size $\kappa$. Numerical analysis is
used to show that the analogous reduction due to random removal of
elements from electrical and elastic networks follow a similar
form. One possible application, namely the use of bone density as
a diagnostic tools for osteoporosis, is discussed.
\end{abstract}
 \maketitle

\section {Introduction}
Networks of electrical or elastic elements are used as models to
study phenomena observed in disordered systems. They include
dielectric breakdown\,\cite{takayasu,dyre,hansen}, metal-insulator
transitions\,\cite{shante}, brittle fracture in disordered
solids\,\cite{sahimi,chung,ray,kantor}, and strength of trabecular
bone\,\cite{gunaratne}. In particular, such models have provided
insights on the critical phenomena\,\cite{herrmann},
scale-invariant disorder\,\cite{herrmann,hansen}, and
size-dependence of the system\,\cite{duxbury}. In this paper, we
will study variations in strength of a class of such systems.

Breakdown processes depend on the size of the longest
defect\,\cite{harlow,leath,finberg} and such extreme value
problems are difficult to analyze\,\cite{herrmann}. However, their
statistical properties are expected to be amenable to analysis;
this is seen to be the case for the class of problems we study
here.

We first consider fused-conducting networks, where the breakdown
currents of each element is assumed to be proportional to its
conductance; i.e., breakdown of a conductor occurs when the
potential difference across it reaches a fixed value. In the
simplest case, conductances along each axis of a hyper-cube are
set equal. In Section II, we calculate how an external current
introduced at a node and removed from an adjacent node is
distributed on the network\,\cite{kirkpatric,bernasconi}. The
Green's Function derived for this case can be used to calculate
the current distribution due to an external field on a network
from which a finite number of conductances are removed (Section
III). This calculation is used to determine how the breaking
current on a network of fused conductances is reduced due to a
fracture of size $\kappa$. The remaining sections of the paper
involve numerical analysis to show how this relationship is
modified in a variety of other electrical and elastic networks.
Section III also provides a discussion of the effects of random
removal of elements from a two dimensional network.

In Section IV, we present analogous results for cubic networks
with randomly chosen conductances. The relationship between the
strength of a network and the fraction of bonds removed is shown
to  have a form similar to relationships found in section III.
Section V shows that the strength of disordered elastic networks
which include elastic and bond-bending energies also follow the
same form.

The motivation for this study is to develop a method to relate the
mean strength of a bone to its mass\,\cite{fung,faulkner}. A
solution of this problem can provide a non-invasive method to
diagnose bone strength and can aid in the management of
osteoporosis. These issues are addressed in the concluding
section.

\section {Green's Functions for Hyper-cubic Network }
Consider an infinite, $d$-dimensional hyper-cubic network where
all conductances along the $\,m^{th}\,$ direction
(${\bf\hat{u}_{m}}$) are assumed to be equal to
$\,\sigma_{m}\,(m=1,2,...,d)\,$. In this section, we will
calculate the current distribution on this network due to a unit
external current introduced at the origin and removed from an
adjacent node, say $\,{\bf{a}}=-{\bf\hat{u}_{1}}\,$. This result
was calculated using a Green's function method by
Kirkpatric\,\cite{kirkpatric,duxbury}, and is reproduced here for
completeness.

Denote the potential at the node
$\,{\bf{n}}=({{n}}_{1},{{n}}_{2},...,{{n}}_{d})\,$, by
$\,V({\bf{n}})\,$ and the current on the conductor joining nodes
${\bf{n}}$ and $\,{\bf{n}}+{\bf\hat{u}_{m}}\,$ by
$\,{\bf{J}}_{m}(\bf{n})\,$; then,
\begin{eqnarray}
{\bf{J}}_{m}({\bf{n}})=\sigma_{m} \left[
V({\bf{n}})-V({\bf{n}}+{{\bf\hat{u}_{m}}})\right]\,.
\end{eqnarray}
The potentials $\,V({\bf{n}})\,$ can be solved using the
Kirchhoff's rules at each node; i.e.,
\begin{eqnarray}\begin{array}{lll}
\sum_{m=1}^{d}&\sigma_{m}&\left\{\left[V({\bf{n}})\,-\,V({\bf{n}}+{\bf{{\hat{u}}}_{m}})\right]
\,\right.\\
&+&\left.\,\,\left[
V({\bf{n}})\,-\,V({\bf{n}}-{\bf{{\hat{u}}}_{m}})\right]\right\}\,=\,\left(\delta_{{\bf{n}},{\bf{0}}}\,-\,\delta_{{\bf{n}},{\bf{a}}}\right)\,,
\end{array}\end{eqnarray}
where the right side is the externally applied current. Eqn. (2)
is easily solved by using the Fourier transform
$\,\hat{V}_{\bf{k}}=\sum_{\bf{n}}e^{-i{\bf{n}}\,{\bf{k}}}V_{\bf{n}}\,,$
which satisfies
\begin{displaymath}
\sum_{m=1}^{d}\,2\,\sigma_{m}\left[1-\cos({\bf\hat{u}_{m}}\cdot{\bf{k}})\right]\,\hat{V}({\bf{k}})=
\left(1-e^{-i{\bf{a}\,k}}\right)\,.
\end{displaymath}
Thus
\begin{eqnarray}
\,\hat{V}({\bf{k}})=
\frac{\left(1-e^{-i{\bf{a\,k}}}\right)}{\sum_{m=1}^{d}\,2\,\sigma_{m}
\left[1-\cos({k_{m}})\right]}\,,
\end{eqnarray}
where $k_{m}={\bf\hat{u}_{m}}\cdot{\bf{k}}$. Hence
\begin{eqnarray}
V({\bf{n}})=\frac{1/2}{\left(2\pi\right)^{d}}\int^{{\bf\pi}}_{-{\bf\pi}}
\,d{\bf{k}}\,\,e^{i{\bf{n}\,k}}\,\,\hat{V}({\bf{k}})\,,
\end{eqnarray}
and
\begin{eqnarray}
\begin{array}{ll}
{\bf{J}}_{m}({\bf{n}})\,=\,\frac{\sigma_{m}/2}{\left(2\pi\right)^{d}}
\,\int^{\bf\pi}_{-\bf\pi}&\,\,d{\bf{k}}\,\,e^{i\bf{n}\bf{k}}\\
\  \\
&\times\frac{\left(1-e^{-i\bf{k_{a}}}-e^{ik_{m}}+e^{i(k_{m}-{\bf{k_{a}}})}\right)}
{\sum_{m\prime=1}^{d}\sigma_{m\prime}\left[1-\cos(k_{m\prime})\right]}\,.
\end{array}
\end{eqnarray}

Next we specialize to networks in two dimensions with equal
conductances ($\sigma_{m}=1$), and define
$\,\alpha(n_{x},n_{y})\equiv\,J_{y}(n_{x},n_{y})\,$ and
$\,\beta(n_{x},n_{y})\equiv\,J_{x}(n_{x},n_{y})\,$, see Figure 1.
The symmetries of the network imply that
\begin{eqnarray}
\begin{array}{ccccc}
\alpha(n_{x},n_{y})&=&\alpha(-n_{x},n_{y})&=&\alpha(n_{x},-n_{y})\,,\\
\beta(n_{x},n_{y})&=&-\beta(-n_{x},n_{y})&=&-\beta(n_{x},-n_{y})\,.
\end{array}
\end{eqnarray}
Combining Eqns. (5) and (6) give
\begin{eqnarray}
\begin{array}{ll}
\alpha(n_{x},n_{y})=\frac{2}{\pi^2}\int^{\pi}_{0}
&\frac{\cos(n_{x}k_{x})\cos(n_{y}k_{y})}
{2-\cos(k_{x})-\cos(k_{y})}\,\sin^{2}(\frac{k_{y}}{2})\,{d\bf{k}},\\
\  \\
\beta(n_{x},n_{y})=\frac{2}{\pi^2}\int^{\pi}_{0}&
\frac{\sin{[(n_{x}+\frac{1}{2})k_{x}]}\sin{[(n_{y}+\frac{1}{2})k_{y}]}}{2-\cos(k_{x})-\cos(k_{y})}\\
&\times\,\sin(\frac{k_{x}}{2})\sin(\frac{k_{y}}{2})\,{d{\bf{k}}}\,.
\end{array}
\end{eqnarray}
Values of $\,\alpha$'s and $\,\beta$'s for several pairs of
$\,(n_{x},n_{y})\,$ are presented in tables I and II, and will be
used for computation in Section III. Notice that
$\,\alpha(0,0)=1/2\,$, as can be easily confirmed using arguments
based on symmetry and superposition\,\cite{kirkpatric}.
\begin{figure}[t]
\centering \vspace*{-0.75cm} \setlength{\abovecaptionskip}{-1.1cm}
\includegraphics[totalheight=3.75in]{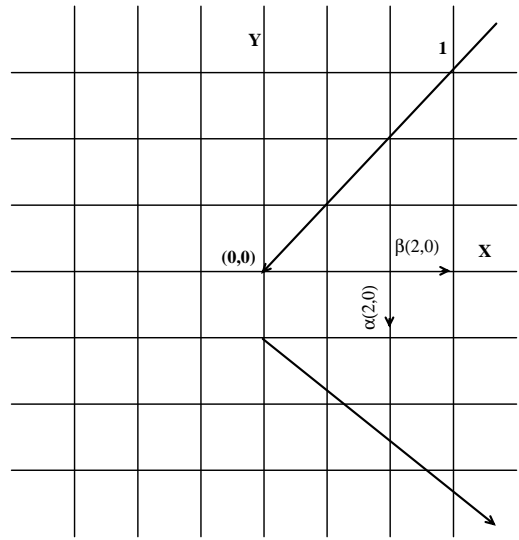}
\caption{\label{fig:epsart} Scheme to compute the current
distribution due to a unit external current introduced at the
origin and removed from an adjacent node.}
\end{figure}
\begin{table}[h]
\caption{\label{tab:table1}Values of
several$\,\,\alpha(n_{x},n_{y})$'s.\,\,In particular
$\,\alpha(0,0)=1/2$, and $\,\alpha(1,0)=1/2(4/\pi-1)$.}
\begin{ruledtabular}
\begin{tabular}{c|ccccccc}
\mbox{$\frac{n_{y}}{n_{x}}$}&0 & 1 & 2 & 3 &4& 5 \\\hline
0&0.50000&0.13662&0.04648&0.02019&0.01080&0.00670\\
1&0.13662&0.00001&0.01455&0.01174&0.00816&0.00571\\
2&0.04648&0.01455&0.00000&0.00404&0.00441&0.00383\\
3&0.02019&0.01174&0.00404&0.00000&0.00162&0.00208\\
4&0.01080&0.00816&0.00441&0.00162&0.00000&0.00080\\
5&0.00670&0.00571&0.00383&0.00208&0.00080&0.00000\\
6&0.00457&0.00413&0.00314&0.00205&0.00113&0.00045
\end{tabular}
\end{ruledtabular}
\end{table}
\begin{table}[h]
\caption{\label{tab:table2} Values of $\,\beta\,$ for the first
pairs of $\,(n_{x},\,n_{y})\,$. }
\begin{ruledtabular}
\begin{tabular}{c|ccccccc}
\mbox{$\frac{n_{y}}{n_{y}}$}&0 & 1 & 2 & 3 &4& 5 \\\hline
0&0.18169&0.04507&0.01315&0.00469&0.00205&0.00106\\
1&0.04507&0.03052&0.01596&0.00826&0.00451&0.00264\\
2&0.01315&0.01596&0.01192&0.00789&0.00510&0.00334\\
3&0.00469&0.00826&0.00789&0.00626&0.00464&0.00337\\
4&0.00205&0.00451&0.00510&0.00464&0.00384&0.00304\\
5&0.00106&0.00264&0.00334&0.00337&0.00304&0.00259\\
6&0.00062&0.00165&0.00225&0.00244&0.00236&0.00214
\end{tabular}
\end{ruledtabular}
\end{table}

The $\alpha$'s and $\beta$'s can also be computed perturbatively
by expanding the denominator as a series in terms of
$\left(\cos{k}_{x}+\cos{k}_{y}\right)/2$. Expansion to fourth
order gives values for $\alpha(n_{x},n_{y})$ and
$\beta(n_{x},n_{y})$ that are accurate to within two percent.

\section {Resistor Networks in Two Dimensions}
Consider next, a two dimensional isotropic
$(\sigma_{x}=\sigma_{y}=1)$ network with an externally applied
unit electric field in the $y-$direction. The currents on this
configuration are
$\,J_{x}(n_{x},n_{y})=0\,,J_{y}(n_{x},n_{y})=1\,,$ see Figure
2(a). In this section , we introduce a method to calculate the
current distribution on the network when a finite number of
conductances are removed. As an application we determine the
enhancement of current on the edges of a ``fracture'' formed by
removing $\kappa$ consecutive conductances.

\subsection {Removal of a Single Conductor}
First, we enumerate changes in the current distribution due to the
removal of the conductor joining the origin to $(0,-1)$, Figure
2(b). The excess current $\delta$ on any remaining element can be
considered to be the redistribution of the current that originally
passed through the link $(0,0)-(0,-1)$, see Figure 2(c). $\delta$
can be evaluated using the values of $\alpha(m,n)$ and
$\beta(m,n)$ by the following procedure. Consider the complete
network, with and external current $(1+J)$ introduced at $(0,0)$
and removed from $(0,-1)$, as shown in Figure 2(d). $(1+J)$ is
chosen so that a current $J$ passes through $(0,0)-(0,-1)$, and
the remainder $(=1)$ passes through the rest of the network; i.e.,
this second part is the solution to the problem illustrated in
Figure 2(c). But from the discussion in section II, the current
passing through $(0,0)-(0,-1)$ is $\alpha(0,0)\times(1+J)$.
\begin{figure}[t]
\centering\vspace*{-0.75cm} \setlength{\abovecaptionskip}{-1.25cm}
\includegraphics[totalheight=4.5in]{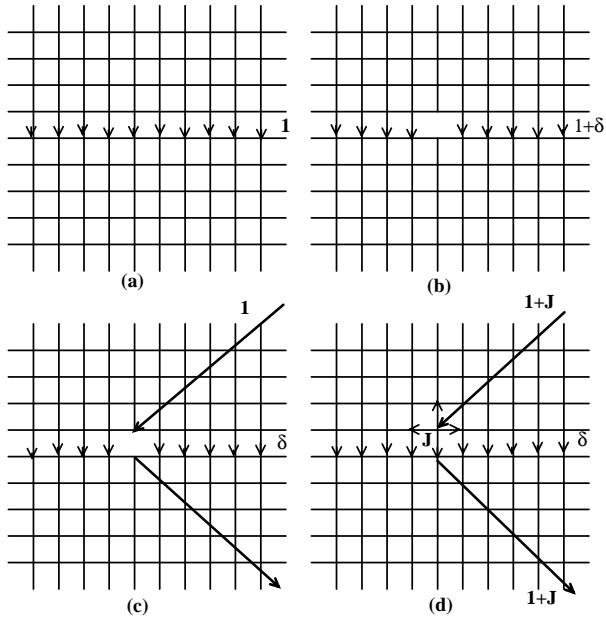}
\caption{\label{fig:epsart} Scheme to compute the current
distribution when a single conductance is removed. In sequence:(a)
an isotropic network with a current flow along the $y-$direction.
In (b) a conductance is removed , and $\delta$ is the excess of
current in any remaining element. Next in (c) , a unit external
current is introduced in order to compute $\delta$. Finally (d)
shows the configuration from which $\delta$ can be evaluated.}
\end{figure}
Hence,
\begin{eqnarray}
{J}=\alpha(0,0)\times\left(1+{J}\right)\,.
\end{eqnarray}
Since $\alpha(0,0)=1/2$, we find that ${J}=1$ and that the changes
of  current in the network are given by,
\begin{eqnarray}
\begin{array}{llll}
\triangle\,J_{y}(n_{x},n_{y})&=(1+J)\alpha(n_{x},n_{y})&=2\,\alpha(n_{x},n_{y})\,\\
\triangle\,J_{x}(n_{x},n_{y})&=(1+J)\beta(n_{x},n_{y})&=2\,\beta(n_{x},n_{y})\,.
\end{array}
\end{eqnarray}
In particular , the current on conductances adjacent to that
removed in Figure 2(b) is
\begin{displaymath}
J_{max}\,=\,1+2\,\alpha(1,0)\,=\,4/\pi\,,
\end{displaymath}
as was given by Duxbury et. al.\,\cite{duxbury}.

As a second illustration, we calculate the currents on the network
shown in Figure 3(a). As in the first example, the problem can be
solved using a complete network on which external currents shown
in Figure 3(b) are applied. Notice that prior to removal of the
conductances there are no current in $AB$ and hence the current
$J_{1}$ that is removed at $B$ is required to pass entirely
through $AB$. The currents $\,J_{1}\,,\,J_{2}\,,\,J_{3}\,$ can
thus be obtained from
\begin{widetext}
\begin{eqnarray}
\begin{array}{ccccccc}
{J}_{1}&=&\mbox{$\alpha(0,0)\,{J}_{1}$}&+&
\mbox{$\beta(0,0)\,(1+{J}_{2})$}&+&\mbox{$\beta(1,0)\,(1+{J}_{3})$}\,,\\
{J}_{2}&=&\mbox{$\beta(0,0)\,{J}_{1}$}&+&
\mbox{$\alpha(0,0)\,(1+{J}_{2})$}&+&\mbox{$\alpha(1,0)\,(1+{J}_{3})$}\,,\\
{J}_{3}&=&\mbox{$\beta(-1,0)\,{J}_{1}$}&+&
\mbox{$\alpha(-1,0)\,(1+{J}_{2})$}&+&\mbox{$\alpha(0,0)\,(1+{J}_{3})$}\,.
\end{array}
\end{eqnarray}
\end{widetext}

Notice that the roles of $\alpha$'s and $\beta$'s are reversed
when computing the distributions due to the external current
$J_{3}$. Solving these equations give $J_{1}=1.50\,,J_{2}=2.38\,,$
and $J_{3}=2.06\,$. Some of the currents on this network are given
in Figure 3(a), and were confirmed through numerical integrations.

Finally we consider the distribution of current due to a
``fracture'' of $\kappa$ consecutive $\sigma_{y}$'s lying along
the $x-$direction, and compute the largest currents (i.e., those
on the conductances forming the edges) on the network. Assume that
the conductances between $(t,0)-(t,-1)$ where $t=1,2,...,\kappa$
are removed. This problem is equivalent to having a complete
network with currents $(J_{t}+1)$ introduced at node $(t,0)$ and
removed at $(t,-1)$ such that currents $J_{t}$ pass through links
$(t,0)-(t,-1)$. These external currents can be obtained by solving
the set of equations
\begin{eqnarray}
J_{t}=\sum_{s=1}^{\kappa}\alpha(s-t,0)\,(J_{s}+1)\,,
\end{eqnarray}
and the current on the conductance $(0,0)-(0,-1)$ is given by
\begin{eqnarray}
J_{max}(\kappa)=1+\sum_{s=1}^{\kappa}\alpha(s,0)\,(J_{s}+1)\,.
\end{eqnarray}
\begin{figure}[b] \centering\vspace*{-0.cm}
\setlength{\abovecaptionskip}{-.4cm}
\includegraphics[totalheight=5.25in]{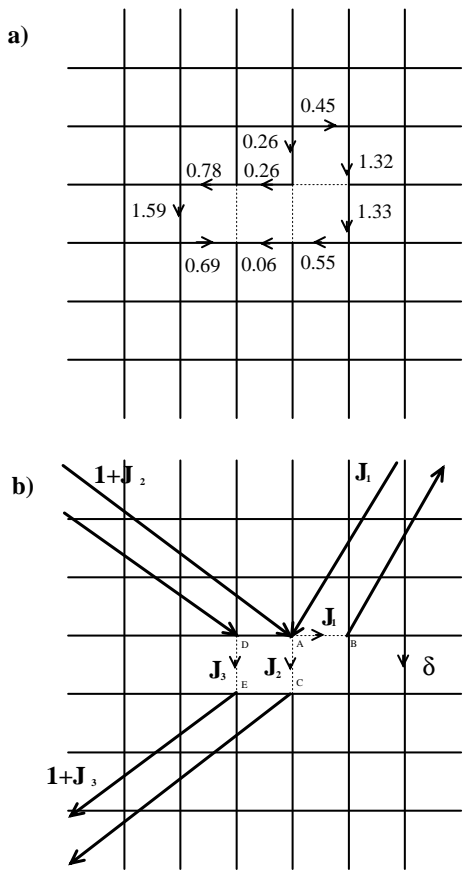}
\caption{\label{fig:epsart}(a) An isotropic network with the three
conductances removed, and current distributions on its boundary.
The scheme to compute $\delta$ is shown in (b).}
\end{figure}
We have solved Eqns. (11) for values of $\kappa$ between $1$ and
$250$, and evaluated the maximum current $J_{max}(\kappa)$ on the
network using Eqn. (12). Their values as a function of $\kappa$
are shown by circles in Figure 4. To a very good approximation,
$J_{max}$ is given by
\begin{eqnarray}
J_{max}(\kappa)\approx\,1+a_{1}\,\kappa^{1/4}+a_{2}\,\kappa^{1/2}\,,
\end{eqnarray}
with $\,a_{1}\approx\,-0.582\,,$ and $\,a_{2}\approx\,0.894\,$;
this expression is shown by the solid line. For $\kappa>>1$, this
approaches the form
$J_{max}(\kappa)=J_{0}\left(1+a\,\kappa^{1/2}\right)\,,$ proposed
in Ref.\cite{duxbury}.
\begin{figure}[t]
\centering \vspace*{-.cm} \setlength{\abovecaptionskip}{-3.25cm}
\includegraphics[totalheight=4.75in]{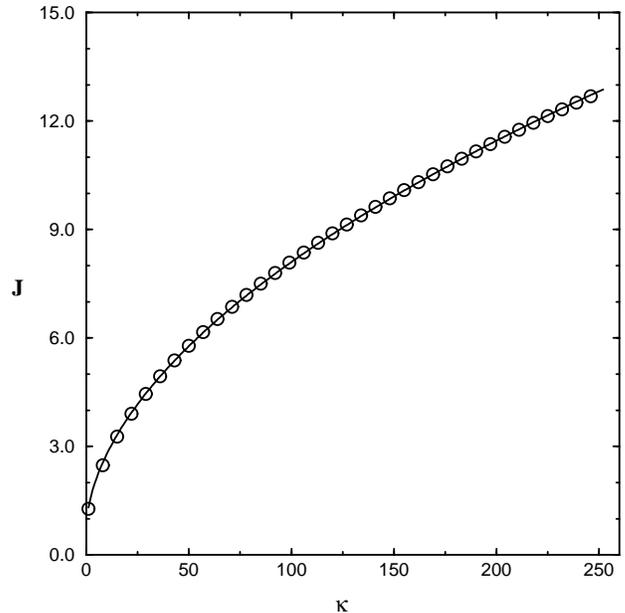}
\caption{\label{fig:epsart} The largest current distribution
resulting from a removal of $\,\kappa\,$ consecutive conductances
lying along the $x-$direction. Results obtained by using Eqns.
(11) and (12) are shown by circles, while the expression (13) is
shown by continuous line.}
\end{figure}
Expressions (11),\,(12) and (13) were confirmed numerically in a
finite network of size $\,L_{x}\times\,L_{y}\,(=100\times\,100)$
centered at the origin. For $\kappa\ge\,45$ effects of the
boundaries become significant and values of $\,J_{max}\,$ near the
boundaries are larger than those given by Eqn. (13).

Next, consider a square network of fused conductances; i.e., if
the current on an element exceeds $I_{0}$, then it will ``burn'',
and its conductance would become zero irreversibly. The complete
network will fuse when the external field is increased to
$E_{0}=I_{max}(0)=I_{0}$, when the current on each element in the
$y-$direction becomes $I_{0}$. The presence of a fracture of
length $\kappa$ will reduce the current $I_{max}(\kappa)$ per link
introduced at the top edge of the network. To calculate
$I_{max}(\kappa)$, notice that the first elements on the network
to burn will be those on the edges of the fracture. As a result
$\kappa$ will increase, and the fracture will propagate in either
direction, until the network is separated into two segments. From
Eqn. (13)
\begin{eqnarray}
\frac{I_{max}(\kappa)}{I_{max}(0)}\approx\frac{1}{1+a_{1}\kappa^{\alpha}+a_{2}\kappa^{2\alpha}}\,,
\end{eqnarray}
with $\,\alpha=1/4\,$. In the remainder of this section an the
next, we will show through numerical analysis, how this form is
modified when resistors are randomly removed from square and cubic
networks. Further, the failure strength of random resistors
networks and disordered elastic networks will be shown to exhibit
a similar form, as long as the networks are not close to the bond
percolation threshold.

We also note that the maximum current $I_{max}$ does not
necessarily correspond to the first fracture of a conductance in
these random cases\,\cite{duxbury,darcangelis}. However, the yield
and ultimate currents have been shown to exhibit similar behavior.

\subsection {Random Removal of Conductances from a Single Layer}
Here we consider a finite network of size $\,L_{x}\times\,L_{y}\,$
and study the reduction of the maximum current, say
$\,I_{max}(\nu)\,$, due to the random removal of a fraction
$\,\nu\,$ of conductances on a {\it\,single layer} $\,y=0\,$. The
length of a largest fracture $\kappa$ on the layer can be
estimated by\,\cite{duxbury}
\begin{eqnarray}
L_{x}\,\nu^{\kappa}\,\left(1-\nu\right)^{2}\,\approx\,1\,,
\end{eqnarray}
where the probability on the left is obtained by requiring the
need for $\kappa$ absent locations (i.e., $\,\nu^{\kappa}\,$)
bounded by two links (i.e., $\,(1-\nu)^{2}\,$). Thus if
$\,(1-\nu)^{2}<<L_{x}\,$
\begin{eqnarray}
 \kappa\,\approx\,\frac{\log(L_{x})}{\log(1/\nu)}.
\end{eqnarray}
Assuming a form similar to Eqn. (14),
\begin{eqnarray}
\frac{I_{max}(\nu)}{I_{max}(0)}=\frac{1}{1+a_{1}\left\{\frac{\log(L_{x})}{\log(1/\nu)}\right\}^{\alpha}
+a_{2}\left\{\frac{\log(L_{x})}{\log(1/\nu)}\right\}^{2\alpha}}\,.
\end{eqnarray}

Figure 5 shows results from computations on a set of 100 networks,
for values of $\,\nu<0.70\,$. In these integrations the boundaries
at the top and bottom were kept at potentials $V_{0}$ and $0$,
respectively. Had the fractures been independent of each other we
would expect $\alpha=1/4$; the fact that $\alpha\approx\,0.37\,$
suggests a strong correlation between
fractures\,\cite{darcangelis}.
\begin{figure}[t]
\centering\setlength{\abovecaptionskip}{-2.25cm}
\includegraphics[totalheight=4.25in]{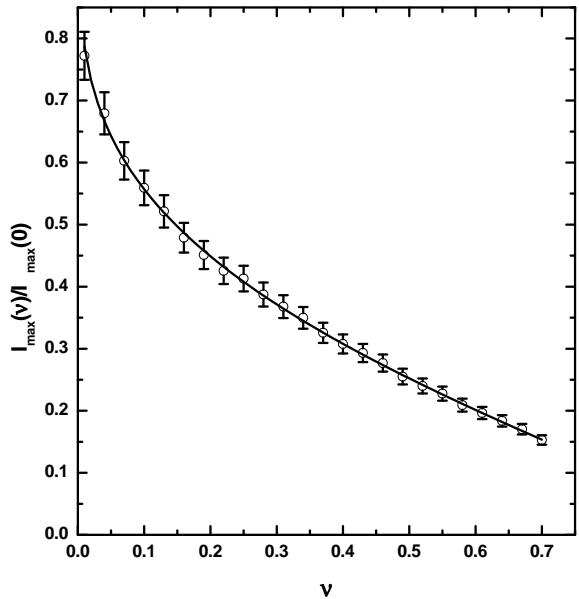} \caption{\label{fig:epsart} The reduction of
$I_{max}(\nu)$ due to the random removal of a fraction $\nu$ of
conductances on a single layer $y=0$ follows Eqn.(17). Here we
show the average and the standard error of a set of 100 networks
for values of $\,\nu<0.70,\,$ with
$\alpha\approx\,0.37,\,a_{1}\approx-0.90$ and
$\,a_{2}\approx\,1.17.$ }
\end{figure}

\subsection {Random Networks in Two Dimensions}
We extend the computations of Section IIB to a network of random
fused conductors from which a fraction $\nu$ of elements have been
removed. The conductances are chosen randomly within a range
$\,[1-\epsilon,1+\epsilon]\,$ and fusing current of each element
is assumed to be proportional to its conductance (i.e., they burn
when the potential difference between the ends exceed a fixed
amount).$\,I_{max}(\nu)\,$ follows a relationship similar to Eqn.
(17) where $\,\alpha\,$ depends on $\,\epsilon\,$.

Consider first the case $\epsilon=0$; i.e. all conductances are
equal. As the fraction $\,\nu\,$ of conductances removed reaches
the bond percolation threshold $\nu_{0}\,(=1/2)$, the network
separates into multiple segments\,\cite{essam,stauffer} , and
hence $I_{max}(\nu_{0})\rightarrow\,0$. Eqn. (17) fails to satisfy
this condition. (Notice that for the quasi-1D case discussed in
Section IIB, $\nu_{0}=1$). We propose a modification
\begin{eqnarray}
\frac{I_{max}(\nu)}{I_{max}(0)}=\frac{1}{1+a_{1}\left\{\frac{\log(N)}{\log(\frac{\nu_{0}}{\nu})}\right\}^{\alpha}+
a_{2}\left\{\frac{\log(N)}{\log(\frac{\nu_{0}}{\nu})}\right\}^{2\alpha}}\,,
\end{eqnarray}
of Eqn. (17) for this case. Here $N=L_{x}\times\,L_{y}$ is the
number of nodes on the network. Figure 6 shows that the mean
values (of $25$ networks of size $100\times\,100$) satisfies Eqn.
(18) with a value of $\alpha\approx\,0.55\,.$ Including randomness
in the conductances changes the value of $\alpha,\,\,a_{1}$ and
$a_{2}$; but de agreement with Eqn.(18) is equally good .
\begin{figure}[t]
\centering\vspace*{-0cm}\setlength{\abovecaptionskip}{-2.25cm}
\includegraphics[totalheight=4.25in]{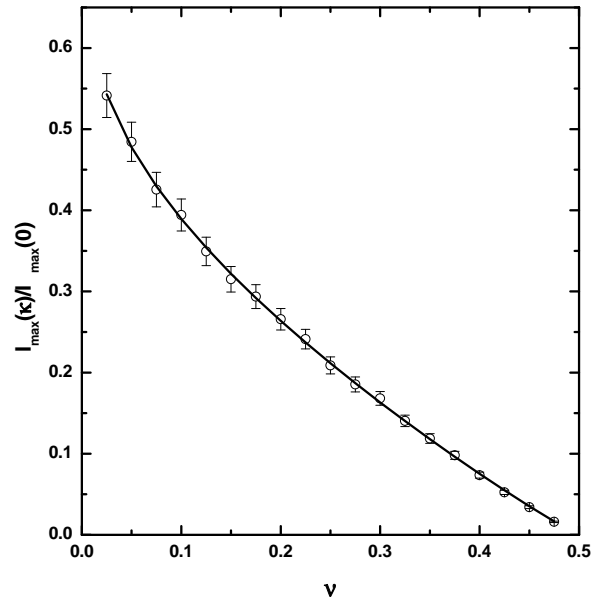}
\caption{\label{fig:epsart} The reduction of the maximum current
as a function of the fraction of conductances removed from a
square network. We considered ensembles of 25 networks with size
$100\times\,100$. Expression (18) is plotted by continuous line
and the average value of
$\alpha\approx\,0.55,\,a_{1}\approx\,0.09\,$and
$\,a_{2}\approx\,0.20.$ }
\end{figure}

\section {Resistor Networks in Three Dimensions}
In this section, we discuss the behavior of the breaking strength
of cubic networks of fused conductances. Green's function
calculations (of Section II) can be used to evaluate the current
distributions via $\alpha(n_{x},n_{y},n_{z})$'s and
$\beta(n_{x},n_{y},n_{z})$'s. Changes in the current distribution
due to the removal of a finite number of conductances can be
evaluated using methods outlined in section IIIA.

\subsection {``Penny-Shaped'' Fracture} We present the calculation
of the current enhancement due to a ``penny-shaped'' fracture in
the $\,Z=0\,$ plane\,\cite{duxbury,harlow}, centered at the
origin. For a given radius $\,\rho\,$ all conductances in the
$z-$direction within a distance $\rho$ of the origin in the
$\,x-y\,$ plane are removed and changes in the current
distribution are computed. This calculation requires evaluating
\begin{eqnarray}
\alpha(n_{x},n_{y},0)=\frac{2}{\pi^{3}}\int^{\pi}_{0}
\frac{\cos(n_{x}k_{x})\cos(n_{y}k_{y})\sin^{2}(k_{z}/2)}{3-\cos(k_{x})-\cos(k_{y})-\cos(k_{z})}d{\bf{k}}\nonumber\\
\end{eqnarray}
Some values of $\alpha(n_{x},n_{y},0)$ are given in TABLE III.
\begin{table}[t]
\caption{\label{tab:table3}Values of $\alpha$'s on the plane
$z=0\,$, for pairs of $\left(n_{x},n_{y}\right)$.}
\begin{ruledtabular}
\begin{tabular}{c|ccccccc}
\mbox{$\frac{n_{x}}{n_{y}}$}& 0 & 1 & 2 & 3&4& 5 \\\hline
0&0.33333&0.06175&0.01392&0.00404&0.00153&0.00073\\
1&0.06175&0.02323&0.00793&0.00298&0.00130&0.00066\\
2&0.01392&0.00793&0.00379&0.00184&0.00095&0.00054\\
3&0.00404&0.00298&0.00184&0.00109&0.00066&0.00041\\
4&0.00153&0.00130&0.00095&0.00066&0.00045&0.00031\\
5&0.00073&0.00066&0.00054&0.00041&0.00031&0.00023\\
6&0.00040&0.00038&0.00033&0.00027&0.00022&0.00017
\end{tabular}
\end{ruledtabular}
\end{table}
\begin{figure}[b]
\centering\vspace*{-.cm}\setlength{\abovecaptionskip}{-3.25cm}
\includegraphics[totalheight=4.75in]{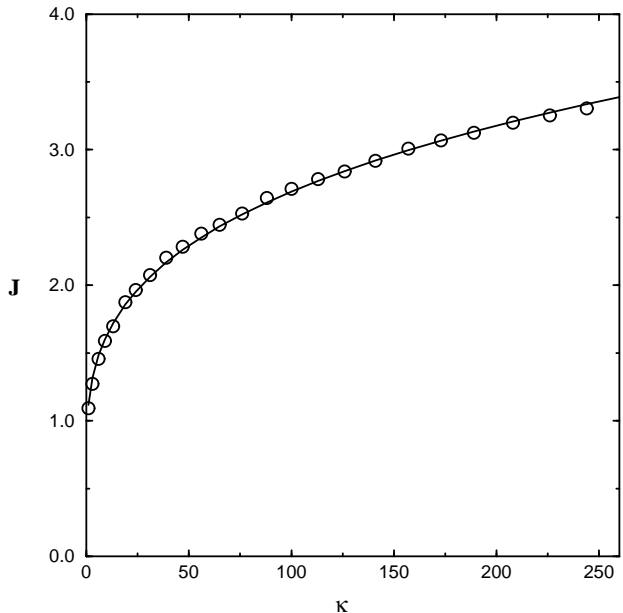}
\caption{\label{fig:epsart}The largest current resulting from a
removal of a ``penny shaped'' fracture of radius $\rho$ lying on
the plane $z=0$. The calculations using Eqn. (19) are shown by
circles, while the expression (20) is shown by the line.}
\end{figure}
The currents on the edge of a fracture can be calculated by
methods similar to those in section III. It is found that the
maximum current for this case is approximately,
\begin{eqnarray}
\,J_{max}(\kappa)\approx\,1+a_{1}\,k^{1/8}+a_{2}\,k^{1/4}\,,
\end{eqnarray}
with $\,a_{1}\approx-0.95\,$  and $\,a_{2}\approx\,1.07\,,$ see
Figure 7. Once again these conclusions were checked in numerical
integrations on cubic networks of conductances.

\subsection {Random Resistor Networks}
In these computations a fraction of conductances from an
$L_{x}\times\,L_{y}\times\,L_{z}$ network was removed randomly. A
potential $\,V_{0}\,$ was imposed at the top surface of the
network, and the bottom surface was grounded. The behavior of the
maximum current $\,I_{max}(\nu)\,$ for increasing $\nu$ is
recorded. As in square networks, $\,I_{max}(\nu)\,$ is seen to
satisfy Eqn. (18), where for cubic networks
$\,\nu_{0}\approx\,0.7508$\,\cite{essam,stauffer}. As before, it
is found that $\alpha$ depends on the level $\,\epsilon\,$ of
disorder.

\section {Elastic Networks}
The elastic network model we consider is constructed by displacing
the nodes of a hyper-cubic network randomly\,\cite{gunaratne}. The
displacements (of less than $\epsilon\,$) are chosen to be small
enough that the topological structure of the network remains
unchanged. Adjacent nodes are joined by struts whose elastic
moduli are chosen randomly from a pre-determined range;
i.e.,$\,k_{e}\in[k_{0}(1-\eta_{e})\,,k_{0}(1+\eta_{e})]\,$.
Bond-bending terms which depend on the variation of the angles
between adjacent bonds on the network are included. The linear
response coefficients are chosen from within a range
$\,\kappa_{b}\in[\kappa_{0}(1-\eta_{b})\,,\kappa_{0}(1+\eta_{b})]\,$.
These networks are degraded by removing a fraction $\nu$ of the
links randomly, and the remaining elastic elements are assumed to
retain their moduli. Thus the total energy of the network is
\begin{displaymath}
U=\sum_{\small{\begin{array}{c}elastic\\elements\end{array}}}\frac{1}{2}\,k_{e}\,\left(\delta\,x\right)^2
+\sum_{\small{\begin{array}{c}bonds\\angles\end{array}}}\frac{1}{2}\,\kappa_{b}\,\left(\delta\theta\right)^2
\end{displaymath}
where $\delta\,x$ is the change of the length of an elastic
element under a given external strain of the network, and
$\delta\theta$ is the variation of a bond-angle from equilibrium.

A strain based criterion is used to implement fracture of elastic
elements and bond angles of the network. Any strut that is
strained beyond a value $\,\gamma\,$ is removed from the network
along with all bond-angles it contributes to. $\gamma\,$ for the
elastic elements are chosen from a Weibull distribution whose
cumulative density is\,\cite{harlow}
\begin{eqnarray}
C(\gamma)=1-exp\left[-\left(\frac{\gamma}{\gamma_{c}}\right)^{m}\right]\,.
\end{eqnarray}
A corresponding criterion is included for fracture of bonds.

A given uniform external strain $\zeta\,$ is imposed through a
sequence of small adiabatic increments. If elastic elements or
bond-angles are removed from the network due to a fracture,
equilibrium is re-calculated prior to increasing $\zeta$. Nodes on
the sides of the network transverse to the external strain are
constrained to move in plane. The force $T(\zeta)$ required to
sustain a strain $\zeta$ is calculated by adding the vertical
forces on the top of the layer. The equilibrium of a network is
calculated by minimizing its energy using the conjugated gradient
method\,\cite{recipes}.

For small values of the compression $\zeta$, the stress-vs-strain
relationship $T(\zeta)$ is linear. Numerical analysis shows that
$T(\zeta)$ becomes nonlinear (at the yield point) following the
first fracture of elastic elements\,\cite{gunaratne}. As $\zeta$
is increased further, the stress reaches a maximum (which, for a
network a fraction $\nu$ of whose elements have been removed, is
denoted $T_{max}(\nu)$) prior to failure.

The analog of Eqn. (18) for an elastic network is
\begin{eqnarray}
\frac{T_{max}(\nu)}{T_{max}(0)}=\frac{1}{1+a_{1}
\left\{\frac{\log{(N)}}{\log{\left(\frac{\nu_{0}}{\nu}\right)}}\right\}^{\alpha}+a_{2}
\left\{\frac{\log{(N)}}{\log{\left(\frac{\nu_{0}}{\nu}\right)}}\right\}^{2\alpha}}\,.
\end{eqnarray}
Figures 8 and 9 show that this expression is satisfied for square
and cubic disordered elastic networks. As in elastic networks,
$\alpha$ depends on the quantities $\epsilon,\,\eta_{e},$ and
$\eta_{b}$ characterizing the level of disorder. It is further
found to depend on the ratio $k_{e}/\kappa_{b}$ of mean linear
response coefficients of elastic and bond-bending  terms.
\begin{figure}[t]
\centering\vspace*{-0cm}\setlength{\abovecaptionskip}{-2.25cm}
\includegraphics[totalheight=4.25in]{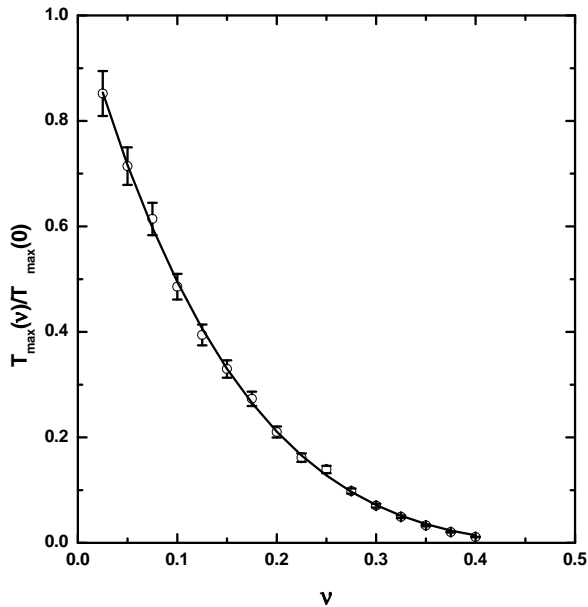}
\caption{\label{fig:epsart}The reduction of $T_{max}(\nu)$ due to
the random removal of a fraction $\nu$ of elastic elements. In
continuous line is shown the average obeying expression (22) of a
square network of size $40\times\,90.\,$ The parameters for the
best fit are
$\alpha\approx\,0.97,\,a_{1}\approx-0.11,\,a_{2}\approx\,0.07.$}
\end{figure}
\begin{figure}[t]
\centering\vspace*{-0cm}\setlength{\abovecaptionskip}{-2.25cm}
\includegraphics[totalheight=4.25in]{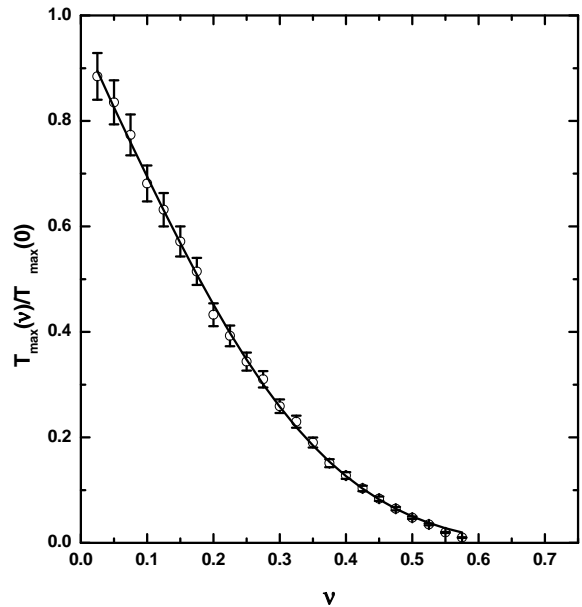}
\caption{\label{fig:epsart} The same as in Figure 8 for a cubic
elastic network of size $20\times\,20\times\,30.\,$For the
nonlinear fit $\alpha\approx\,1.14,\,a_{1}\approx-0.011,$ and
$a_{2}\approx\,0.015.$}
\end{figure}

\section {Discussions and Conclusions}
The motivation for the work reported here is the need to identify
measures that can be used as non-invasive diagnostic tools to
estimate the strength of bone. Large bones consist of an outer
compact shaft and an inner porous region (trabecular architecture)
whose structure is reminiscent of a network of disordered cubic
networks\,\cite{gunaratne}. Due to loss of fracture
toughness\,\cite{finberg} of the outer shaft, the trabecular
architecture becomes the principal load carrier in older adults.
Reduction in sex-steroid hormones (estrogen and testosterone)
effect bone turnover, leading to degradation of the porous bone.
Beyond this stage, it is important to monitor the loss of bone
strength at regular intervals, preferably with non-invasive
diagnosis tools.

The principal mechanism for loss of trabecular mass is through
removal of individual struts, due to minor trauma. Fracture
toughness and thickness of the remaining trabecular bone do no
change appreciably. Mechanical studies on ex-vivo bone samples
have shown that trabecular networks from patients with a broad
range of ages fracture at a fixed level of strain\,\cite{hogan},
even though the corresponding fracture stresses exhibit large
variations. These observations motivated the fracture criteria
used in our models.

Eqn. (18) is the main result of the paper. Its form is justified
by the calculations on a network of fused conductors presented in
Section IIIA. The numerical analysis of random electrical networks
and disordered elastic networks are used to confirm that the
strength bears the same relationship to the level of elements
removed from the network. The index $\alpha$ corresponding to
random removal of elements is larger that in the case of a single
fracture (=1/4), indicating that correlations between fractures is
important to determine the strength of a
network\,\cite{darcangelis}.

It is important to test if samples of trabecular bone satisfy Eqn.
(18). Previous studies on bone samples have suggested a nonlinear
relationship between the breaking strength $(T)$ and the effective
density $(\rho)$ of trabecular bone. It was found heuristically
that $T\approx\rho^{-\beta}$, where $\beta\approx 2.6$. If it is
assumed that the density of bone material remains intact and that
the fractional reduction of effective density is proportional to
$\nu$, then the relationship Eqn. (18) can be tested using this
data. Such a comparison will be done in future work.

The authors would like to thank M.Marder, G. Reiter and
S.Wimalawansa for discussions. This work is partially funded by
the Office of Naval Research, the National Science Foundation and
the ICSC - World Laboratory.


\begin{thebibliography}{apssamp}
\bibitem{takayasu}H. Takayasu, Phys. Rev. Lett. {\bf\,54},\,
 1099\,(1985)
\bibitem{dyre}J.C. Dyre. and Th. Schoder B, Rev. of Modern Phys.
 {\bf\,72},\,873\,(2000)
\bibitem{hansen}A. Hansen, E.L. Hinrichsen and S. Roux, Phys. Rev. B
{\bf\,43},\,665\,(1991) and in Phys. Rev. Lett. {\bf{66}},\,
2476\,(1991)
\bibitem{shante}V.K.S. Shante and S. Kirkpatric, Adv.
Phys. {\bf\,20},\,325\,(1971)
\bibitem{sahimi} M. Sahimi and J.D. Goddard, Phys. Rev.
B {\bf\,32},\,1869\,(1985)
\bibitem{chung}J.W. Chung, A. Ross, J. Th. De Hosson  and E van der Giessen, Phys. Rev. B {\bf\,54},\,094\,(1996-I)
\bibitem{ray}P. Ray and B.K. Chakrabarti, Phys. Rev. B {\bf\,38},\,715\,(1988)
\bibitem{kantor}Y. Kantor and I. Webman, Phys. Rev. Lett. {\bf\,52},\,1891\,(1984)
\bibitem{gunaratne}G.H. Gunaratne, C.S. Rajapakse, K.E. Bassler, K.K. Mohanty and S.J. Wimalawansa, Phys. Rev. Lett.
{\bf\,88},\,68101\,(2002)
\bibitem{herrmann}H.J. Herrmann and S. Roux, Phys. Rev. B{\bf{39}},\,637\,(1989)
\bibitem{duxbury}P.M. Duxbury, P.D. Beale and P.L. Leath, Phys. Rev. Lett {\bf\,57},\,1052\,(1986) and in Phys. Rev. B
{\bf\,36},\,367\,(1986)
\bibitem{harlow}D.G. Harlow and S.L. Phoenix, Int. J. Frac.
{\bf\,17},\,601\,(1981)
\bibitem{leath}P.L. Leath and P.M. Duxbury, Phys. Rev. B {\bf\,49},\,14905\,(1994)
\bibitem{finberg}J. Finberg and M. Marder, Phys. Rep. {\bf\,313},\, 1\,(1999)
\bibitem{kirkpatric}S. Kirkpatric, Rev. of Modern Phys. {\bf\,
45},\, 574\,(1973)
\bibitem{bernasconi}J. Bernasconi, Phys. Rev. B {\bf\, 9},\, 4575\,(1974)
\bibitem{fung}Y.C. Fung, ``Biomechanics: Mechanical Properties of
Living Tissue'', Springer-Verlag, New York,\,1993
\bibitem{faulkner}K.G. Faulkner, J. Bone. Miner. Res. {\bf\,15},\,183\,(2000)
\bibitem{darcangelis}L. De Arcangelis and H.J. Herrmann, Phys.
Rev. B {\bf{39}},\,2678\,(1989)
\bibitem{essam}J.W. Esam, Rep. Prog. Phys. {\bf{43}},\,53\,(1980)
\bibitem{stauffer}D. Stauffer, `` Introduction to Percolation
Theory'', Taylor \& Francis, London Philadelphia,\,1985
\bibitem{recipes}W.H. Press et.al., ``Numerical Recipes-The Art of
Scientific Computing'', Cambridge University Press,
Cambridge,\,1988
\bibitem{hogan}H.A. Hogan, S.P. Ruhman and N.H. Sampsom, J. Bone
Miner. Res. {\bf{15}},\,284\,(2000)
\bibitem{morgan}E.F. Morgan and T.M. Keaveny, J. Biomech.,
{\bf{34}},\,569\,(2001)
\end{thebibliography}
\end{document}